\documentstyle[12pt]{article}

\begin{document}

\newcommand{\beq}{\begin{equation}}
\newcommand{\eeq}{\end{equation}}
\newcommand{\barr}{\begin{eqnarray}}
\newcommand{\earr}{\end{eqnarray}}

\newcommand{\andy}[1]{ }

\def\a{\alpha_0} \def\da{\delta\alpha}
\def\Da{D_\alpha}
\def\rz{\mbox{\boldmath $y$}_0}
\def\h{\widehat}
\def\t{\widetilde}
\def\cH{{\cal H}}



\author{Hiromichi NAKAZATO$^{(1)}$\thanks{Permanent address:
Department of Physics, University of the Ryukyus, Okinawa 903-01,
Japan} $\;$
 \& Saverio PASCAZIO$^{(2)}$ \\
        $^{(1)}$Institut f\"ur Theoretische Physik,
Universit\"at Wien  \\
A-1090 Wien, Austria  \\
        $^{(2)}$Dipartimento di Fisica,
Universit\^^ {a} di Bari \\
and Istituto Nazionale di Fisica Nucleare, Sezione di Bari \\
I-70126  Bari, Italy}

\title{Macroscopic limit of a solvable dynamical model }

\date{}

\maketitle

\begin{abstract}
The interaction between an ultrarelativistic particle and a
linear array made up
of $N$ two-level systems (^^ ^^ AgBr" molecules) is
studied by making use of a modified version of the Coleman-Hepp
Hamiltonian.
Energy-exchange processes between the particle and the molecules
are properly taken into account, and the evolution of the total system is
calculated exactly both when the array is initially in the ground state
and in a thermal state.

In the macroscopic limit ($N \rightarrow \infty$),
the system remains solvable and leads to interesting
connections with the Jaynes-Cummings model, that describes the interaction
of a particle with a maser.

The visibility of the interference pattern produced by the two
branch waves of the particle is computed, and
the conditions under which the spin array in the $N \rightarrow \infty$
limit behaves as a ^^ ^^ detector" are investigated.
The behavior of the visibility yields good insights into the issue of
quantum measurements:
It is found that, in the thermodynamical limit,
a superselection-rule space appears in the description
of the (macroscopic) apparatus.

In general, an initial thermal state of the ^^ ^^ detector"
provokes a more substantial loss of quantum coherence than an initial
ground state. It is argued that a system decoheres more as the temperature
of the detector increases. The problem of ^^ ^^ imperfect measurements"
is also shortly discussed.
\end{abstract}

\vspace*{.5cm}
PACS: 03.80.+r; 03.65.Fd; 42.52.+x

\newpage

\tableofcontents

\newpage


\setcounter{equation}{0}
\section{Introduction  }
\label{sec-introd}
\andy{intro}

Quantum mechanics is considered to be a fundamental theory of nature,
due to its successful predictions in many practical applications.
Nevertheless, we still lack a complete understanding of its
interpretative postulates, in particular on the so-called
quantum measurement problem \cite{von}.
There is not even unanimous consensus about the definition of the
problem itself, and in fact there have been long discussions in order to
understand whether a quantum mechanical
measurement process can be analyzed within the quantum  mechanical
formalism \cite{Zurek}.

von Neumann's projection rules \cite{von}
are very useful computational tools, but the presence
of an external ^^ ^^ classical" measuring apparatus is required in
order to provoke the ^^ ^^ wave-function collapse". We feel that this is not
satisfactory, because a measuring system is made up of
elementary constituents that must be treated quantum mechanically,
and is therefore a quantum mechanical object itself.
On the other hand, the ^^ ^^ classical" nature of the macroscopic
measuring system should be properly taken into account, because
we know that the above-mentioned von Neumann's
rules work well, in practical calculations.

In this paper we shall give a concrete example of interaction between an
elementary quantum system $Q$ and a model detector $D$.
Notice that if we want to treat the $Q+D$ system quantum mechanically,
we must consider the quantum mechanical structure of both systems,
and this is highly nontrivial if one of the two systems is made up
of many elementary constituents, because we are forced to consider
the interaction between the object particle $Q$ and every single
elementary constituent of $D$.

In order to study the interaction between an object particle and
a detection system in the above-mentioned sense,
solvable models are very helpful: Not only they give
good insights into physics, but they also provide us with a
better understanding of the complicated phenomena involved.
In this respect, a model Hamiltonian proposed by Hepp \cite{AgBr}
is very well known: It describes the interaction between an ultrarelativistic
particle and an ensemble of two-level systems, and is usually referred
to as ^^ ^^ AgBr" or Coleman-Hepp model.
Due to its relative simplicity, the model has received
considerable attention in the past, and has played an important role
in the literature on the measurement problem
\cite{BellHepp} \cite{NamFound} \cite{KMK}
\cite{NP3} \cite{NaPa2}.
Another interesting solvable model which describes the interaction
between a two-level system and the electromagnetic field in a cavity
(maser)
was proposed by
Jaynes and Cummings some years ago \cite{JC}.

Our purpose is twofold.
First, by making use of a modified version recently proposed \cite{NaPa3}
for the AgBr Hamiltonian, we will study the interaction between the
particle and the detector when the latter is initially in a thermal state.
This situation is more realistic than the usual one, in which
the detector is initially taken to be
in the ground state, because $D$ is
macroscopic and cannot be completely isolated from its environment.
We emphasize that it would be impossible to study the case of a
thermal detector starting from the original Coleman-Hepp model,
due to the absence of a free Hamiltonian for $D$.
The introduction of the latter will also enable us to compute several
physically relevant quantities, such as the energy ^^ ^^ stored" in $D$
as a result of the interaction, its fluctuation and their ratio.

Second, we shall consider a macroscopic limit for the AgBr model.
We shall see that there is a connection between this limit and
the Jaynes-Cummings model.
The link can be seen only in our modified version,
that is able to take into account
energy-exchange processes between $Q$ and $D$.

We will realize that
the macroscopic limit of a detection system is
extremely important from the point of view of
quantum measurements: Indeed,
the visibility of the interference pattern
can be exactly computed for the case of finite number $N$ of
elementary constituents of $D$,
and its behaviour in the $N\to\infty$ limit is very interesting.
It will be seen that a macroscopic system does not necessarily behave
as a ^^ ^^ detector", unless other important conditions are met.

This paper is organized as follows.
We review the original Coleman-Hepp model in Sec.~2,
and introduce the modified version in Sec.~3, where the
case of a detector initially in a thermal state is
also considered. In Sec.~4, we compute the $N\to\infty$ limit of some
interesting physical quantities and of the scattering matrix.
A slightly modified version of the Jaynes-Cummings model is displayed
in Sec.~5, and the relevant evolutions are considered.
We will see that the latter model yields the same results
obtained in the macroscopic limit from the AgBr case.
In this limit and under certain
conditions, the two Hamiltonians are shown to be identical in Sec.~6.
The correspondence will be pushed further in Sec.~7, where the
problem of quantum measurements will be considered,
in particular in the light of the appearance of unitary-inequivalent
representations in the many-Hilbert-space theory
proposed by Machida and Namiki
\cite{MN}. Section~8 contains additional considerations concerning
this issue and shortly touches upon the concept of imperfect measurements.

\setcounter{equation}{0}
\section{Review of the original AgBr Hamiltonian }
\label{sec-agbrgen}
\andy{agbrgen}

Let us start off by introducing the
Coleman-Hepp or AgBr Hamiltonian, and by reviewing the
main results obtained by different authors in the past
\cite{AgBr} \cite{BellHepp} \cite{NamFound} \cite{KMK}
\cite{NP3} \cite{NaPa2}. Even though the content
of the present section is not original, new light will be thrown on those
results that are most important from the ^^ ^^ macroscopic"
point of view to be analyzed in the present paper.

The AgBr Hamiltonian
describes the interaction between an
ultrarelativistic particle $Q$ and a one-dimensional $N$-spin array
($D$-system).
One can think, for instance, of a linear emulsion  of AgBr
molecules, the {\em down} state  corresponding  to  the  undivided
molecule, and the {\em up} state corresponding to the  dissociated
molecule (Ag and Br atoms).
The particle and each molecule interact via a spin-flipping local potential.

The total Hamiltonian for the $Q+D$ system is
\andy{tothamorig}
\barr
H & = & H_{Q} + H^{\prime(0)},
\label{eq:tothamorig}
\earr
where $H_{Q}$ is the free Hamiltonian of the particle and $H^{\prime(0)}$
the interaction Hamiltonian.  These are explicitly written as
\andy{Horig}
\barr
H_{Q} & = & c \h{p},    \nonumber \\
H^{\prime(0)} & = & \sum_{n=1}^{N}  V(\h{x}- x_n)  \sigma_{1}^{(n)} \ ,
\label{eq:Horig}
\earr
where $\h{p}$ is the momentum  of
the particle, $\h{x}$ its position, $V$ is a real  potential,  $x_n
\; (n=1,...,N)$ are the positions of the scatterers in the  array
and $\sigma_{1}^{(n)}$ is the
Pauli matrix acting on the $n$th
site.

This Hamiltonian is a nice model of a typical measurement process and can
be solved exactly. Let us sketch rapidly
the main results by making use of
generalized coherent states \cite{KMK}.
The evolution operator in the interaction picture
\andy{evol}
\beq
U(t,t') = e^{iH_{Q}t/\hbar} e^{-iH(t-t')/\hbar}
              e^{-iH_{Q}t'/\hbar}      \label{eq:evol}
\eeq
can be computed exactly as
\andy{solut}
\beq
U(t,t') = \exp \left( -\frac{i}{\hbar} \int_{t'}^{t}
H_{I}^{\prime(0)}(t'') dt'' \right),      \label{eq:solut}
\eeq
where the interaction Hamiltonian in the interaction picture
is given by
\andy{HINT}
\beq
H_{I}^{\prime(0)}(t)  = e^{iH_{Q}t/\hbar} \, H^{\prime(0)}
e^{-iH_{Q}t/\hbar}
     = \sum_{n=1}^{N} V(\h{x} + ct- x_n)
  \sigma_{1}^{(n)} .
\label{eq:HINT}
\eeq
A straightforward calculation yields the
following $S$-matrix \andy{Smatr}
\beq
S^{[N]} = \lim_{\stackrel{{\scriptstyle t\rightarrow
\infty}}{t'\rightarrow -\infty}} U(t,t')  =
\exp \left( -i\frac{V_{0} \delta}{\hbar c}
\sum_{n=1}^{N} \sigma_1^{(n)} \right) =
\prod_{n=1}^{N} S_{(n)} \ ,
\label{eq:Smatr}
\eeq
where \andy{prodo}
\beq
S_{(n)} = \exp \left( -i\frac{V_{0} \delta}{\hbar c}
\sigma_1^{(n)} \right) =
\cos \left( \frac{V_{0} \delta}{\hbar c} \right)
-i \sigma_1^{(n)}
\sin \left( \frac{V_{0} \delta}{\hbar c} \right),
  \label{eq:prodo}
\eeq
and $V_{0} \delta \equiv \int_{-\infty}^{\infty} V(x)dx$.
This allows us to define the ^^ ^^ spin-flip" probability, i.e.
the probability of dissociating one AgBr molecule, as
\andy{sfprob}
\beq
q = \sin^{2} \left( \frac{V_{0} \delta}{\hbar c} \right) .
 \label{eq:sfprob}
\eeq

Notice that the Hamiltonian $H$ is invariant under exchange of scatterers in
the array. Therefore, if we call ${\cal P}_N$ the group of permutations on
$\{ 1, \ldots, N \}$, we can restrict our attention to the
${\cal P}_N$-invariant
sector ${\cal H}_N$ of the bigger Hilbert space ${\cal H}_{\{N\}}$ of the
$N$ scatterers. The former is generated by the symmetrized states
$\vert j>_N, \; j=1, \ldots, N$,
where $j$ is the number of dissociated molecules,
while the latter by the vectors $\vert \{j\}>_N$, representing states in
which $j$ {\em particular} molecules are dissociated.
The two types of vectors are related to each other via the formula
\andy{legame}
\beq
\vert j>_N = {N\choose j}^{-1/2} \sum_{\{j\}} \vert \{j\}>_N ,
\label{eq:legame}
\eeq
where the summation $\sum_{\{j\}}$ is over the permutations. Incidentally,
observe that $\dim {\cal H}_{\{N\}} = 2^N$,
while $\dim {\cal H}_{N} = N+1$. In the following, we shall concentrate
our analysis on the {\em symmetrized} case, and give only a few comments
for the
other case. The symmetrization will become a delicate problem in the $N
\rightarrow \infty$ limit, to be tackled in the following sections.

The $S$-matrix can alternatively be written as
\andy{alter}
\beq
S^{[N]} =  \exp \left( -i\frac{V_{0} \delta }{\hbar c}
 N \Sigma_1^{(N)} \right) ,
\label{eq:alter}
\eeq
where \andy{Sigmaj}
\beq
\Sigma^{(N)}_{j}  =  \frac{1}{N}
 \sum_{n=1}^{N}
\sigma_{j}^{(n)}, \qquad  j=1,2,3      \label{eq:Sigmaj}
\eeq
is the average spin.
Observe that \andy{comm}
\beq
\left[ N \Sigma^{(N)}_{i}, N
\Sigma^{(N)}_{\ell}  \right]  =  2i
 N \Sigma^{(N)}_{k},     \label{eq:comm}
\eeq
with ${i,\ell,k}$ any even permutation of ${1,2,3}$, so that
the operators $N \Sigma_j$ form a unitary representation
of $SU(2)$.
Moreover, by defining  \andy{sigpm}
\begin{equation}
\Sigma^{(N)}_\pm  =  \frac{1}{2} \left(
 \Sigma^{(N)}_1 \pm  i
\Sigma^{(N)}_2 \right),         \label{eq:sigpm}
\end{equation}
one gets the algebra
\andy{newalg}
\barr
\left[ N \Sigma^{(N)}_-,N \Sigma^{(N)}_+ \right] & = &
-N \Sigma^{(N)}_3, \nonumber \\
\left[ N \Sigma^{(N)}_-,-N \Sigma^{(N)}_3 \right] & = &
-2N \Sigma^{(N)}_-,          \label{eq:newalg}  \\
\left[ N \Sigma^{(N)}_+,-N \Sigma^{(N)}_3 \right] & = &
+2N \Sigma^{(N)}_+.         \nonumber
\earr
The initial $D$-state is taken to be the ground state
$\vert  0>_N$ ($N$ spins down), and we shall first
consider the situation in which the initial $Q$-state is a plane wave.
The evolution is easily computed
from eq.(\ref{eq:alter})
by observing that
\andy{KMKorigrel}
\barr
N \Sigma^{(N)}_+ \vert n>_N & = &
    \sqrt{(N-n)(n+1)} \; \vert n+1>_N,
\nonumber \\
N \Sigma^{(N)}_- \vert n>_N & = &
    \sqrt{(N-n+1)n} \; \vert n-1>_N,
         \label{eq:KMKorigrel}  \\
N \Sigma^{(N)}_3 \vert n>_N & = &
    (2n-N) \; \vert n>_N ,
\nonumber
\earr
and by making use of the formula
\cite{Bogol} \andy{rel}
\barr
e^{-i \alpha \cdot N \Sigma^{(N)}_1} & = &
        e^{\tanh(-i\alpha)\cdot N\Sigma^{(N)}_+}
        e^{-\ln\cosh(-i\alpha)\cdot N\Sigma^{(N)}_3}
        e^{\tanh(-i\alpha)\cdot N\Sigma^{(N)}_-}
\nonumber \\
        & = &
        e^{-i \tan\alpha \cdot N \Sigma^{(N)}_+}
        e^{-\ln\cos\alpha \cdot N \Sigma^{(N)}_3}
        e^{-i \tan\alpha \cdot N \Sigma^{(N)}_-}.
     \label{eq:rel}
\earr
The result is
\andy{Svac}
\barr
S^{[N]} \vert p, 0>_N & = & \sum_{j=0}^N  \sum_{\{j\}}
(-i \sqrt{q} \, )^j \left( \sqrt{1-q} \, \right)^{N-j}
      \vert p, \{j\}>_N  \nonumber \\
   & = &  \sum_{j=0}^N {N\choose j}^{1/2}
  \left( -i\sqrt{q} \, \right)^j \left( \sqrt{1-q} \right)^{N-j}
 \vert p, j>_N,
\label{eq:Svac}
\earr
where we have used the notation $\vert p,\{j\}>_N = \vert p>\vert \{j\}>_N, \;
\vert p,j>_N = \vert p>\vert j>_N$.
The far right hand side in eq.(\ref{eq:Svac}) is a generalized coherent state
\cite{KMK}.

In a typical interference experiment a
divider splits an incoming wave function $\psi$ into two branch
waves $\psi_1$ and $\psi_2$, so that the initial state of the $Q+D$ system is
\andy{psi}
\begin{equation}
\Psi_I = \left( \psi_1 + \psi_2 \right) \vert 0>_N \ ,         \label{eq:psi}
\end{equation}
where $\vert  \psi_i > = \int dp_i c(p_i) \vert p_i> \; \; (i=1,2)$ are
one-dimensional wave packets, normalized to unity. Assume that only
$\psi_2$ interacts with $D$.
The final state of the total system is
\andy{after}
\begin{equation}
\Psi_F = \vert \psi_1> \vert 0>_N  + S^{[N]}
  \vert \psi_2> \vert 0>_N \ ,
\label{eq:after}
\end{equation}
and after recombination of the two branch waves
the probability of observing the particle is \andy{square}
\begin{equation}
P = \vert \Psi_F \vert^2 =
\vert \psi_1 \vert^2+\vert\psi_2\vert^2
      +2\Re\Bigl[ \psi_1^*\psi_{2}
      \;_N \! < 0 \vert S^{[N]} \vert 0>_N \Bigr].
\label{eq:square}
\end{equation}
Interference is observed when a phase shifter is inserted in one of the two
paths (neutron-interferometer type), or when the two branch waves originating
from the slits are forwarded to a distant screen (Young-interferometer
type). In both cases, the visibility is readily calculated
by eqs.(\ref{eq:Svac}) and (\ref{eq:square}) as
\andy{vis}
\begin{equation}
{\cal V} = \frac{P_{\rm MAX} - P_{\rm min}}{P_{\rm MAX} + P_{\rm min}}  =
 \;_N \! <0\vert  S^{[N]} \vert 0>_N  =
 \left(1 - q \right)^{N/2}.
     \label{eq:vis}
\end{equation}
Equations~(\ref{eq:alter}), (\ref{eq:Svac})
and (\ref{eq:vis}) are the main results of the above analysis.
Observe that the result is exact and holds true for every value of $N$.
The $N \rightarrow \infty$ limit is a somewhat delicate problem, and
will be one of the main objectives of the present study.

Notice that, as was to be expected, for finite $q \neq 0$, the
interference pattern disappears in the $N$-infinity
limit. This is essentially the case considered by Hepp \cite{AgBr}
and Bell \cite{BellHepp}.
On the other hand, the limit
$N \rightarrow \infty, \; qN= \overline{n}=$
finite \cite{NP3} is more interesting:
In this case, the visibility becomes
\andy{vislim}
\beq
{\cal V} \stackrel{N \rightarrow \infty, \, qN= {\rm
finite}}{\longrightarrow} e^{-q N/2} =
e^{-\overline{n}/2}.       \label{eq:vislim}
\eeq
Note that  $qN=\overline{n}$  represents  the  average  number  of
excited molecules, so that interference gradually disappears as
$\overline{n}$ increases. This is in contrast with the
^^ ^^ sudden" disappearance of
interference in the finite $q \neq 0$ case.

It remains to be stressed that the Hamiltonian $H$ can be shown
\cite{NaPa2}  to  be equivalent to the
one  studied  in  Ref. \cite{Cini}, if we restrict
our attention to the Hilbert space
${\cal H}_N$.

\setcounter{equation}{0}
\section{The modified AgBr Hamiltonian }
\label{sec-modifagbr}
\andy{modifagbr}

The previous results are very interesting, but
we should remark that the above interaction Hamiltonian does not take
into account the possibility of energy exchange between the particle and the
spin system:
Both systems never lose (or gain) energy as a consequence of the interaction.
According to Ref.~\cite{Araki},
a measuring apparatus that is not affected by the interaction simply
acts as a ^^ ^^ decomposer", i.e.~a device that is only able to perform a
spectral decomposition. In order to obtain a change of the apparatus state
reflecting the state of the measured system one must, in general, modify
the Hamiltonian of the total system.
In the Coleman-Hepp case, even though the state of the spin array changes
and the total energy of the $Q+D$ system is conserved,
the energy levels of the spin system are completely neglected.
This is not satisfactory, if we want to regard the spin system as a
detecting device, because we are implicitly assuming to be able to
distinguish {\em energetically}
different states of the array:  On the other hand,
this can be made only via a free Hamiltonian of the spin system, which is
absent in the above description.

This situation  can be improved \cite{NaPa3}
by taking into account both the energy levels of the $D$-system and the
energy transfer between the $Q$ and $D$ systems:
The free Hamiltonian of the spin array is added, and
an appropriate operator is introduced into the interaction Hamiltonian.
These modifications make the model more consistent and realistic.
Remarkably, the model remains solvable
if a ^^ ^^ resonance condition" is met.

The total Hamiltonian for the $Q+D$ system becomes
\andy{totham}
\barr
H & = & H_{0} + H',  \nonumber \\
 & & \qquad H_0 = H_{Q} + H_{D},
\label{eq:totham}
\earr
where the free Hamiltonians of the particle and
of the detector, $H_{Q}$ and $H_{D}$, and the modified interaction
Hamiltonian $H'$ are written as
\andy{H}
\barr
H_{Q} & = & c \h{p},    \nonumber \\
H_{D}  &  =  &  \frac{1}{2}  \hbar  \omega
  \sum_{n=1}^{N}  \left( 1+\sigma_{3}^{(n)} \right) , \nonumber \\
H' & = & \sum_{n=1}^{N}  V(\h{x}- x_n)
  \sigma_{1}^{(n)} \exp \left( i \frac{\omega}{c}
  \sigma_{3}^{(n)} \h{x} \right)  \label{eq:H} \\
    & = & \sum_{n=1}^{N} V(\h{x}- x_n)
  \left[ \sigma_{+}^{(n)} \exp \left( -i \frac{\omega}{c}
  \h{x} \right) + \sigma_{-}^{(n)} \exp \left( + i \frac{\omega}{c}
  \h{x} \right) \right]. \nonumber
\earr
Notice that the
energy difference between the two states of the molecule is $\hbar
\omega$, and that the previous Hamiltonian (eq.(\ref{eq:Horig}))
is reobtained in the $\omega \rightarrow 0$ limit.

Observe that, in contrast with every previous analysis \cite{AgBr}
\cite{BellHepp} \cite{NamFound} \cite{KMK} \cite{NP3} \cite{NaPa2},
we are not neglecting the free energy of
the scatterers, represented by $H_{D}$,
and are taking into account the energy exchange between the
$Q$-particle and the spin system: This is accomplished by the above
interaction Hamiltonian, whose action can be decomposed in the following way
\andy{resoncond}
\barr
H'_{(n)} \vert p,\downarrow_{(n)}> & = & V(\h{x} - x_n)
     \vert p - \frac{\hbar \omega}{c},
     \uparrow_{(n)}> ,
\nonumber \\
H'_{(n)} \vert p,\uparrow_{(n)}> & = & V(\h{x} - x_n)
     \vert p + \frac{\hbar \omega}{c},
     \downarrow_{(n)}> ,
\label{eq:resoncond}
\earr
where $H'_{(n)}$ is the $H'$-term  acting on the $n$th site,
$\vert p,\downarrow_{(n)}>$ represents a state in which the $Q$-particle
has momentum $p$ and the $n$th
molecule is undivided (spin down), and analogously for
the other cases. We understand
from eq.(\ref{eq:resoncond}) that the interaction
Hamiltonian $H'$ satisfies a ^^ ^^ resonance condition", because the energy
acquired or lost by the $Q$-particle in every single interaction
matches exactly the energy gap between the two spin states
(i.e.~the energy required to provoke one spin flip).

The analysis of the previous section is readily extended to the present
case. The $S$-matrix stems from the product of factors \cite{NaPa3}
\andy{prodobis}
\barr
S_{(n)} & = & \exp \left( -i\frac{V_{0} \delta}{\hbar c}
\mbox{\boldmath $\sigma^{(n)} \cdot u$} \right) =
\cos \left( \frac{V_{0} \delta}{\hbar c}\right)
-i \mbox{\boldmath $\sigma^{(n)} \cdot u$}
\sin \left( \frac{V_{0} \delta}{\hbar c}\right),
  \label{eq:prodobis}  \\
  & & \qquad \mbox{\boldmath $u$}
 = \left( \cos \left( \frac{\omega}{c} x \right),
 \sin \left( \frac{\omega}{c} x \right), 0 \right) ,
  \nonumber
\earr
and is computed as
\andy{Smatrbis}
\beq
S^{[N]} = \prod_{n=1}^{N} S_{(n)} =
\exp \left( -i\frac{V_{0} \delta}{\hbar c} \sum_{n=1}^N
\mbox{\boldmath $\sigma^{(n)} \cdot u$} \right) =
\exp \left( -i\frac{V_{0} \delta}{\hbar c} N
\mbox{\boldmath $\Sigma^{(N)} \cdot u$} \right) .
\label{eq:Smatrbis}
\eeq
(Compare with eqs.(\ref{eq:Smatr}) and~(\ref{eq:alter}),
and observe that the spin-flip probability
$q$ is the same.)
We shall now compute the evolution of the total system in two interesting
cases.

\subsection{Initial ground state }
\label{sec-groundagbr}
\andy{groundagbr}

If we take, as in the previous section, the ground state $\vert 0>_N$
as the initial $D$-state, the evolution of the total system
is easily calculated as \andy{Svacbis}
\barr
S^{[N]} \vert p, 0>_N & = & \sum_{j=0}^N  \sum_{\{j\}}  (-i
\sqrt{q} \, )^j \left( \sqrt{1-q} \, \right)^{N-j}
\vert p -j\frac{\hbar \omega}{c}, \{j\}>_N  \nonumber \\
 & = & \sum_{j=0}^N {N\choose j}^{1/2}
 \left( -i\sqrt{q} \right)^j \left( \sqrt{1-q} \right)^{N-j}
 \vert p -j\frac{\hbar \omega}{c}, j>_N .
\label{eq:Svacbis}
\earr
Once again, we obtain the value
\andy{visbis}
\begin{equation}
{\cal V} = \frac{P_{\rm MAX} - P_{\rm min}}{P_{\rm MAX} + P_{\rm min}}  =
 \;_N \! <0\vert  S^{[N]} \vert 0>_N  =
 \left(1 - q \right)^{N/2}
     \label{eq:visbis}
\end{equation}
for the visibility of the interference pattern.
Notice that, for the sake of simplicity, we are suppressing the
dependence on the ^^ ^^ screen coordinate", in
the second equality of eq.(\ref{eq:visbis}) (see Appendix A).
In the following, we shall always suppress the $Q$-states
unless confusion may arise.

It is interesting to calculate the energy ^^ ^^ stored" in the
array after the interaction with the particle.
It is computed as
\andy{avener}
\beq
<H_D>_F = \,_N\!<0\vert  S^{[N]\dagger} H_D S^{[N]} \vert 0>_N
    = qN \, \hbar \omega ,
\label{eq:avener}
\eeq
where $F$ stands for final state
and the $Q$-particle states are suppressed.
The fluctuation around the average is
\andy{flucavener}
\beq
<\delta H_D>_F = \sqrt{< \left( H_D - <H_D>_F \right)^2 >_F}
  = \sqrt{pqN} \, \hbar \omega ,
\label{eq:flucavener}
\eeq
where $p=1-q$, and their ratio is given by
\andy{flucratio}
\beq
\frac{<\delta H_D>_F}{<H_D>_F} = \sqrt{\frac{p}{qN}} .
\label{eq:flucratio}
\eeq
We stress that the above results
(\ref{eq:avener})-(\ref{eq:flucratio})
are new, and could not be calculated starting from the
original Coleman-Hepp Hamiltonian
(\ref{eq:tothamorig}) and (\ref{eq:Horig}),
due to the absence of the free Hamiltonian
$H_D$.

The limit $N \rightarrow \infty, \; qN=\overline{n} < \infty$
is most interesting and will be discussed in the next section.
We shall see that such a limit can be consistently taken only for the
^^ ^^ modified" AgBr Hamiltonian introduced in this section.


\subsection{Initial thermal state }
\label{sec-thermo}
\andy{thermo}

In the previous subsection we have considered the interaction between a
$Q$-particle and a spin array $D$ when the latter
is initially in the ground state. This situation is not completely
satisfactory, from the physical
point of view: Indeed, our spin array is a caricature of a detector, and is
therefore a macroscopic object. A more realistic description of $D$ should
therefore take into account such macroscopic quantities as volume, temperature
and so on. Let us now consider the case in which the detector is initially
in a thermal state, characterized by the density matrix
\andy{thinit}
\beq
\rho_{\rm th} = \frac{1}{Z} \exp \left[ -\beta \frac{\hbar \omega}{2}
  \sum_{n=1}^{N}  \left( 1+\sigma_{3}^{(n)} \right) \right] ,
\label{eq:thinit}
\eeq
where $Z$ is the partition function and
$\beta= 1/k \theta$, $\theta$ being the temperature.
As previously stated, we restrict ourselves to the
symmetrized space ${\cal H}_N$, so that
the identity is written as
${\bf 1} = \sum^N_{j=0} \vert j>_{NN}<j\vert $, and
\andy{repth}
\beq
\rho_{\rm th} = \frac{1}{Z} \sum^N_{j=0} \exp \left[ - j \beta \hbar \omega
  \right] \vert j>_{NN}<j\vert .
\label{eq:repth}
\eeq
The condition $\mbox{Tr} \rho_{\rm th}=1$ yields
\andy{Zfunc}
\beq
Z= \frac{1- e^{- \beta \hbar \omega (N+1)}}{1-e^{- \beta \hbar \omega}}.
\label{eq:Zfunc}
\eeq
(Incidentally, notice that in the unsymmetrized space
${\cal H}_{\{N\}}$ there would appear
different expressions for the above two quantities.)

We are implicitly assuming that the interaction between $Q$ and $D$
takes place when our detector is in contact with a thermal
{\em reservoir}, at temperature $\theta$. Obviously, after the interaction,
$D$ will thermalize again, returning eventually to its initial
state $\rho_{\rm th}$, so that no trace of the passage of the
$Q$-particle will be left.
This situation is not very interesting, from our point of view,
because we are just investigating under which conditions the spin array
responds to the interaction with the $Q$-particle, detecting its passage.
Only in such a case the $D$-system can be considered as a ^^ ^^ detector".
In the following analysis we shall assume that the coupling between $D$
and the thermal {\em reservoir} is very weak compared to that between
$D$ and $Q$, so that the state of $D$ immediately after the interaction with
$Q$
can be considered, to a very good approximation, as the final state.
Alternatively, we can assume that the interaction between $Q$ and $D$ is
much quicker than between $D$ and the {\em reservoir}, so that the
thermalization process of $D$ after its interaction with $Q$
requires a much longer time.

The initial $D$-state is characterized by the quantities
\andy{inarray}
\barr
<H_D>_I^{\rm th} & = & \mbox{Tr} \left( H_D \rho_{\rm th}
          \right) = \hbar \omega \left( \frac{e^{-\beta \hbar \omega}}{1 -
          e^{-\beta \hbar \omega}} -
            \frac{(N+1)e^{-(N+1)\beta \hbar \omega}}{1 -
          e^{-(N+1)\beta \hbar \omega}} \right) ,   \nonumber \\
<\delta H_D >_I^{\rm th} & = & \hbar \omega \left(
            \frac{e^{-\beta \hbar \omega}}{\left( 1 -
          e^{-\beta \hbar \omega} \right)^2 } -
            \frac{(N+1)^2 e^{-(N+1)\beta \hbar \omega}}{\left( 1 -
          e^{-(N+1)\beta \hbar \omega}\right)^2 } \right)^{1/2} ,
\label{eq:inarray}
\earr
where $I$ stands for initial state.

The evolution of the $Q+D$ system
can be computed explicitly, but the final state, expressed in
terms of density matrices, does not have a simple expression
due to the presence of the $Q$-particle states.
We shall see, in Sec.~\ref{sec-games},
how it is possible to devise a formal expedient
in order to get rid of the $Q$-states.
Here, we just calculate the value of the physically interesting
quantities in the following way:
{}From eq.(\ref{eq:prodobis}) we get
\andy{alterS}
\beq
S_{(n)} = \exp \left[ -i \varpi
\mbox{\boldmath $\sigma^{(n)} \cdot u$} \right] =
  \exp \left[ -i \varpi \left( \sigma^{(n)}_+ e^{-i \frac{\omega}{c}\h{x}}
  + \sigma^{(n)}_- e^{i \frac{\omega}{c}\h{x}}
 \right) \right] ,
  \label{eq:alterS}
\eeq
where we have written $\varpi = V_{0} \delta / \hbar c$.
(Notice that if we assume a small spin-flip probability
$q$ (of order $N^{-1}$), we get
$q = \sin^{2} \varpi \simeq \varpi^2$.)
It is then easy to compute
\andy{partev}
\barr
S_{(n)} \sigma_3^{(n)} S^{\dagger}_{(n)}
  & = & \sigma_3^{(n)} \cos 2\varpi +
        \left(\sigma^{(n)}_1 \sin \frac{\omega}{c}\h{x}
         - \sigma^{(n)}_2 \cos \frac{\omega}{c}\h{x} \right) \sin 2 \varpi
    \nonumber \\
  & = & \sigma_3^{(n)} \cos 2\varpi +
        i \left(\sigma^{(n)}_+ e^{-i \frac{\omega}{c}\h{x}}
         - \sigma^{(n)}_- e^{i \frac{\omega}{c}\h{x}} \right) \sin 2 \varpi
    \nonumber \\
  & \equiv & \sigma_3^{\star(n)} ,
\label{eq:partev}
\earr
so that
\andy{Devol}
\barr
S^{[N]} H_D S^{[N]\dagger}
  & = & \frac{\hbar \omega}{2}
  S^{[N]} \sum_{n=1}^{N}  \left( 1+\sigma_{3}^{(n)} \right) S^{[N]\dagger}
            \nonumber \\
  & = & \frac{\hbar \omega}{2}
  \sum_{n=1}^{N}  \left( 1+\sigma_{3}^{\star(n)} \right) .
\label{eq:Devol}
\earr
It is then straightforward, if lenghty, to prove that
\andy{thenerF}
\barr
<H_D>_F^{\rm th} & = & \mbox{Tr} \left( H_D \rho_{\rm th}^F
          \right)   \nonumber \\
& = & \mbox{Tr} \left( H_D S^{[N]}
          \rho_{\rm th} S^{[N]\dagger}
          \right)   \nonumber \\
        & = & \mbox{Tr} \left( S^{[N]\dagger} H_D S^{[N]} \rho_{\rm th}
          \right)   \nonumber \\
        & = & \cos 2 \varpi <H_D>_I^{\rm th}
          + \frac{N}{2} \hbar \omega (1 - \cos 2 \varpi) ,
\label{eq:thenerF}
\earr
where $F$ stands for final state.

Analogously, we get
\andy{thenerfluc}
\barr
& &
< \! \delta H_D \! >_F^{\rm th} \; =  \nonumber \\
& &  \quad = \Biggl[ \left( \frac{\hbar \omega}{2} \right)^2 \!
        \left[ \left( 4 \cos^2 2 \varpi - 2 \sin^2 2 \varpi \right) \!
         \left( 2 \gamma^2_1 + \gamma_1 - (N+1)
         (N+1 + 2 \gamma_1)  \gamma_{N+1} \right) \right.
   \nonumber \\ & &
\quad  + \left. \left( 4 N \cos 2 \varpi  (1 - \cos 2 \varpi) + 2 N \sin^2 2
        \varpi \right) \left( \gamma_1 - (N+1) \gamma_{N+1} \right) \right.
   \nonumber \\ & & \left.
\quad  + N^2  (1 - \cos 2 \varpi)^2 + N \sin^2 2 \varpi  \right]
      - \left( <H_D>_F^{\rm th} \right)^2 \Biggr]^{1/2},
\label{eq:thenerfluc}
\earr
where $\gamma_m = \exp (-m \beta\hbar\omega)/
(1-\exp (-m \beta\hbar\omega))$.

The calculation for the visibility is more involved and is explained in
Appendix~A. The final result is
\andy{visN}
\barr
{\cal V}^{\rm th} & = & \mbox{Tr} \left( \rho_{\rm th} S^{[N]} \right)
 =  \frac{e^{\beta \hbar \omega} \cos^{N+2} \varpi }{Z(t_+ - t_-)}
     \left( \frac{1}{t_-^{N+1}} - \frac{1}{t_+^{N+1}} \right) ,
\label{eq:visN}      \\
& & t_\pm = \frac{1}{2} \left[
       \cos^2 \varpi \left( 1+e^{\beta \hbar \omega} \right)
       \pm \sqrt{\cos^4 \varpi \left( 1+e^{\beta \hbar \omega} \right)^2
         - 4 \cos^2 \varpi e^{\beta \hbar \omega}}\, \right] .
 \nonumber
\earr
Obviously, we recover the results of the
previous subsection for $\theta = 0 \; (\beta = \infty)$.

Once again, we realize the advantage of keeping the free Hamiltonian
$H_D$. If one started from the original Hamiltonian
(\ref{eq:tothamorig}) and (\ref{eq:Horig}),
one would not be able to discuss the
temperature dependence of the physically interesting
quantities (see eqs.(\ref{eq:inarray}), (\ref{eq:thenerF}),
(\ref{eq:thenerfluc}) and (\ref{eq:visN})):
Indeed, if there are no energy differences between different spin
configurations, the $D$-system, if it is to be represented by a mixture,
is always described by the density matrix of a completely random
ensemble \cite{Sakurai}, irrespectively of the temperature, and any discussion
about the temperature dependence would be meaningless.

\setcounter{equation}{0}
\section{The $N \rightarrow \infty$ limit  }
\label{sec-JaCu}
\andy{JaCu}

One of the main purposes of the present investigation is
to study the thermodynamical limit of the
modified AgBr model,
introduced in the previous section.
This will be done by keeping
the quantity $qN$ always finite.
The physical meaning of
this limit is appealing: It corresponds
to admitting that the number of dissociated molecules
$\overline{n} = q N $ is finite. Alternatively, one can say
that the energy $\overline{n}\, \hbar \omega = q N \hbar \omega$
exchanged between the particle and the detector is kept finite, even though the
number of elementary constituents of $D$ becomes very large.

\subsection{Interesting physical quantities}
\label{sec-groundmaser}
\andy{groundmaser}

Let us now evaluate the physical quantities calculated in the previous section
in the $N \rightarrow \infty$, $\overline{n} = qN =$ finite
limit.

If the initial $D$-state is the ground state $\vert 0>_N$ we obtain, from
(\ref{eq:avener}), (\ref{eq:flucavener}), (\ref{eq:flucratio}) and
(\ref{eq:visbis}), respectively,
\andy{alllim}
\barr
<H_D>_F & \rightarrow  & \overline{n} \, \hbar \omega ,    \nonumber \\
<\delta H_D>_F & \rightarrow  & \sqrt{\overline{n}} \,
     \hbar \omega ,        \nonumber \\
\frac{<\delta H_D>_F}{<H_D>_F} & \rightarrow  &
     \frac{1}{\sqrt{\overline{n}}},
      \label{eq:alllim}  \\
{\cal V} \quad & \rightarrow & e^{- \overline{n} /2}. \nonumber
\earr
Notice that the simple
$N \rightarrow \infty$ limit, with finite
$q \neq 0$, yields only divergent or
vanishing quantities.

On the other hand, if we start from the thermal $D$-state we get,
from eq.(\ref{eq:inarray}),
\andy{thlim1}
\barr
<H_D>_I^{\rm th} & \rightarrow & \hbar \omega n_{\rm th} ,
                    \nonumber \\
<\delta H_D>_I^{\rm th} & \rightarrow  & \hbar \omega \, n_{\rm th}
            e^{\beta\hbar\omega/2} ,
 \label{eq:thlim1} \\
\frac{<\delta H_D>_I^{\rm th}}{<H_D>_I^{\rm th}} & \rightarrow  &
            e^{\beta\hbar\omega/2},
\nonumber
\earr
where $n_{\rm th} = e^{-\beta\hbar\omega} /
\left( 1-e^{-\beta\hbar\omega} \right)$
is the number of excited (up) spins in the initial thermal state,
and from eqs.(\ref{eq:thenerF})-(\ref{eq:visN}),
\andy{thlim}
\barr
<H_D>_F^{\rm th} & \rightarrow & \hbar
            \omega \left( n_{\rm th} + \overline{n} \right) ,
                    \nonumber \\
<\delta H_D>_F^{\rm th} & \rightarrow  & \hbar \omega \, n_{\rm th}
            \sqrt{ 1 + \left( 2 \overline{n} + 1 \right)
            \left( e^{\beta\hbar\omega}-1 \right)
            + \overline{n} \left( e^{\beta\hbar\omega}-1 \right)^2 } ,
         \nonumber \\
\frac{<\delta H_D>_F^{\rm th}}{<H_D>_F^{\rm th}} & \rightarrow  &
            \frac{\sqrt{1 + \left( 2 \overline{n} + 1 \right)
            \left( e^{\beta\hbar\omega}-1 \right)
            + \overline{n} \left(e^{\beta\hbar\omega}-1 \right)^2 }}{ 1 +
            \overline{n} \left(e^{\beta\hbar\omega}-1 \right) },
 \label{eq:thlim} \\
{\cal V}^{\rm th}  & \rightarrow &
 \exp \left[ - \left( n_{\rm th} + \frac{1}{2} \right)
           \overline{n} \right] .
 \nonumber
\earr
Obviously, we recover the previous results
(eq.(\ref{eq:alllim})) for $\theta = 0 \; (\beta = \infty)$.
In particular,
\andy{thfluclim}
\beq
\frac{<\delta H_D>_F^{\rm th}}{<H_D>_F^{\rm th}} \left\{
  \begin{array}{ll}
   \stackrel{\beta=0}{\longrightarrow} & 1  \\
   \stackrel{\beta=\infty}{\longrightarrow} & 1/\sqrt{\overline{n}} .
  \end{array}
\right.
\label{eq:thfluclim}
\eeq

\subsection{The scattering matrix }
\label{sec-Smatrlim}
\andy{Smatrlim}

Let us now turn our attention to the $N \rightarrow \infty$ limit of the
scattering matrix $S^{[N]}$ (eq.(\ref{eq:Smatrbis})). Observe
that the $S$-matrix
can be rewritten as \andy{newSmatr}
\barr
S^{[N]} & = &
\exp \left[ -i\frac{V_{0} \delta}{\hbar c} \sum_{n=1}^N
\mbox{\boldmath $\sigma^{(n)} \cdot u$} \right]
    \nonumber \\
   & = & \exp \left[ -i
  \frac{V_{0} \delta}{\hbar c} \sum_{n=1}^{N}
  \left( \sigma_{+}^{(n)} \exp \left[ -i \frac{\omega}{c}
  \h{x} \right] + \sigma_{-}^{(n)} \exp \left[ + i \frac{\omega}{c}
  \h{x} \right] \right) \right]  \nonumber \\
   & = & \exp \left[ -i
         \frac{V_{0} \delta}{\hbar c} \sqrt{N} \left(
         \exp \left[ -i \frac{\omega}{c} \h{x} \right]
         \frac{1}{\sqrt{N}} \sum_{n=1}^{N} \sigma_{+}^{(n)}
         + \exp \left[ + i \frac{\omega}{c} \h{x} \right]
         \frac{1}{\sqrt{N}} \sum_{n=1}^{N} \sigma_{-}^{(n)}
         \right) \right] . \nonumber \\
\label{eq:newSmatr}
\earr
Consider now that the condition $qN=$ finite, with
$\sqrt{q} \simeq V_0 \delta / \hbar c$,
implies that the quantity $(V_0 \delta / \hbar c) \sqrt{N} \equiv
u_0 \delta / \hbar c$
behaves ^^ ^^ well" in the $N \rightarrow \infty$ limit, i.e.~it does
neither diverge, nor vanish.
(We assume, for simplicity and without loss of generality,
that $\delta$ is the same quantity used in
Sec.~\ref{sec-agbrgen}.)
On the other hand, the operators
$N^{-1/2} \sum_{n=1}^{N} \sigma_{\pm}^{(n)}$ and
$(1/2) \sum_{n=1}^{N} \left( 1+\sigma_{3}^{(n)} \right)$
(which is nothing but the free Hamiltonian of the detector in
eq.(\ref{eq:H})) obey, in the $N \rightarrow \infty$ limit,
the standard boson commutation relations for
$a, a^\dagger$ and ${\cal N} = a^\dagger a$ \cite{contrac}.
This is shown explicitly in
Appendix~B. Summing up, we can identify
\andy{sumup}
\barr
\frac{1}{\sqrt{N}} \sum_{n=1}^{N} \sigma_{+}^{(n)}
 = \sqrt{N} \Sigma_{+}^{(N)}
   & \stackrel{N \rightarrow \infty}{\longrightarrow} & a^\dagger, \nonumber \\
\frac{1}{\sqrt{N}} \sum_{n=1}^{N} \sigma_{-}^{(n)}
 = \sqrt{N} \Sigma_{-}^{(N)}
& \stackrel{N \rightarrow \infty}{\longrightarrow} & a,
 \label{eq:sumup}    \\
\frac{1}{2} \sum_{n=1}^{N} \left( 1 + \sigma_{3}^{(n)} \right)
 = \frac{N}{2} \left( {\bf 1}^{(N)} + \Sigma_{3}^{(N)} \right)
   & \stackrel{N \rightarrow \infty}{\longrightarrow} & {\cal N}
   \equiv a^\dagger a ,
    \nonumber
\earr
so that the $S$-matrix becomes
\andy{SlimJC}
\beq
S^{[N]}
 \rightarrow S = \exp \left[ -i\frac{u_{0} \delta}{\hbar c}
         \left( a^\dagger \exp \left[ -i \frac{\omega}{c} \h{x} \right]
         + a \exp \left[ i \frac{\omega}{c} \h{x} \right]
         \right) \right] .
  \label{eq:SlimJC}
\eeq
The connection with a ^^ ^^ maser" system is obvious, and will be
made more precise in the next section.


\setcounter{equation}{0}
\section{The maser system }
\label{sec-masersys}
\andy{masersys}

Let us clarify the connection between the modified AgBr and
the maser systems. First we consider the case in which the $D$-system is
an electromagnetic field in a cavity (maser).
We keep the free Hamiltonian $H_Q = c \h{p}$ for the $Q$-system,
so that the total Hamiltonian is given by
\andy{tothamJC}
\barr
H^{\rm JC} & = & H_{0}^{\rm JC} + H^{\prime {\rm JC}},  \nonumber \\
 & & \qquad H_0^{\rm JC} = H_{Q} + H_{D}^{\rm JC},
\label{eq:tothamJC}
\earr
where
\andy{HJC}
\barr
H_Q  & = & c \h{p} , \nonumber \\
H_{D}^{\rm JC} & = & \hbar \omega {\cal N} = \hbar \omega
     a^\dagger a, \nonumber \\
H^{\prime {\rm JC}} & = & u(\h{x})
  \left[ a^\dagger \exp \left( -i \frac{\omega}{c} \h{x} \right)
         + a \exp \left( i \frac{\omega}{c} \h{x} \right)  \right].
\label{eq:HJC}
\earr
Here, we wrote JC in order to stress the resemblance with
the so-called Jaynes-Cummings \cite{JC} Hamiltonian, that describes the
interaction between a two-level system and the electromagnetic field in a
cavity. The JC Hamiltonian differs from the present one
only because it contains terms of the type $\tau_\pm$, instead of
$\exp \left(\pm i \frac{\omega}{c} \h{x} \right)$, $\tau_\pm$ being the
raising/lowering operators for a two-level system.
In the case we are considering, the $Q$-particle has a continuous spectrum,
and can exchange an arbitrary number of quanta of energy $\hbar \omega$.
Clearly, this difference is not
important for our analysis, one of the purposes of which is to understand the
behaviour of the spin array in the $N \rightarrow \infty$ limit.

We solve the interaction between the $Q$-particle and the
$D$-system (maser).
The interaction Hamiltonian in the interaction picture is
\andy{HIJC}
\beq
H^{\prime {\rm JC}}_I = u(\h{x}+ct)
  \left[ a^\dagger \exp \left( -i \frac{\omega}{c} \h{x} \right)
         + a \exp \left( i \frac{\omega}{c} \h{x} \right)  \right],
\label{eq:HIJC}
\eeq
and yields the $S$-matrix
\andy{SJC}
\beq
S = \exp \left[ -i\frac{u_{0} \delta}{\hbar c}
         \left( a^\dagger \exp \left[ -i \frac{\omega}{c} \h{x} \right]
         + a \exp \left[ i \frac{\omega}{c} \h{x} \right]
                       \right) \right],
  \label{eq:SJC}
\eeq
where $\int u(x) dx = u_0 \delta$, $\delta$ being
the same quantity used in eq.(\ref{eq:newSmatr}).
Notice that the $S$-matrix obtained here is exactly the same as that derived in
the $N \rightarrow \infty$ limit for the modified AgBr model
(\ref{eq:SlimJC}).
We will see below that the results obtained in
Secs.~\ref{sec-groundagbr} and~\ref{sec-thermo} can be extended to the
^^ ^^ JC" case in a fully consistent way. The analogy between the two cases
will be pushed much further in
Sec.~\ref{sec-games}.
We first calculate the physically interesting quantities for the JC case
in this section, and then put forward a correspondence between the modified
AgBr and JC Hamiltonians in
Sec.~\ref{sec-Hlim}.

\subsection{Initial ground state}
\label{sec-maserground}
\andy{maserground}

Let us first assume
the initial state to be $\vert p,0>=\vert p>\vert 0>$,
where $\vert 0>$ is the
ground state of the maser cavity.
The evolution is
\andy{SvacJC}
\beq
S \vert p, 0> = e^{- \overline{\kappa}/2}
   \sum_{j=0}^\infty \frac{(-i \sqrt{\overline{\kappa}} \, )^j}{\sqrt{j!}}
   \vert p - j \frac{\hbar \omega}{c}, j>,
  \qquad \overline{\kappa} = \left( \frac{u_0 \delta}{\hbar c} \right)^2 ,
\label{eq:SvacJC}
\eeq
where $\vert p_j, j> = \vert p_j>\vert j>$,
$\vert j>$ being the number state of the cavity.
By observing that
\andy{newnil}
\beq
e^{i \alpha (a^\dagger + a)} a e^{-i \alpha (a^\dagger + a)} = a -i \alpha,
\label{eq:newnil}
\eeq
we easily obtain
\andy{JCgro}
\barr
<H^{\rm JC}_D>_F & = &
  \overline{\kappa} \, \hbar \omega ,    \nonumber \\
<\delta H^{\rm JC}_D>_F & = & \sqrt{\overline{\kappa}} \,
     \hbar \omega ,        \nonumber \\
\frac{<\delta H^{\rm JC}_D>_F}{<H^{\rm JC}_D>_F} & = &
     \frac{1}{\sqrt{\overline{\kappa}}},     \label{eq:JCgro} \\
{\cal V}^{\rm JC} \quad & = &
   e^{- \overline{\kappa} /2}, \nonumber
\earr
where $F$ denotes the final state and the matrix elements of the
$Q$-particle states are trivially computed.
As was to be expected, the above equations
allow us to interpret $\overline{\kappa}$ as the average number of
boson excitations in the cavity.
We can see the perfect correspondence between eqs.(\ref{eq:alllim})
and~(\ref{eq:JCgro}), if we identify
\andy{nk}
\beq
\overline{n} \Leftrightarrow \overline{\kappa}.
\label{eq:nk}
\eeq
Incidentally, notice that, if we neglect altogether the $Q$-particle states,
the generalized coherent state of eq.(\ref{eq:Svacbis})
becomes, in the $N \rightarrow \infty, \, qN < \infty$
limit, the Glauber coherent state
of eq.(\ref{eq:SvacJC}). We shall come back to this
point in Sec.~\ref{sec-games}.

The analogy between the two cases, i.e. between
the macroscopic limit ($N \rightarrow
\infty$) of the $N$-spin system and the maser system,
has thus been estabilished when the ground state is chosen as the initial
$D$-state. What happens if we choose a thermal state as initial state?
We shall analyze this case in the next subsection.


\subsection{Initial thermal state }
\label{sec-JCth}
\andy{JCth }

The thermal state of the cavity can be written
\andy{thinitJC}
\beq
\rho_{\rm th}^{\rm JC} = \frac{1}{\cal Z} \exp \left[ -\beta \hbar \omega
                a^\dagger a \right] ,
\label{eq:thinitJC}
\eeq
where ${\cal Z}$ is the partition function and
$\beta= 1/k \theta$, $\theta$ being the temperature.
In the Fock space ${\cal H}$ the identity is
${\bf 1} = \sum^\infty_{j=0} \vert j><j\vert $, so that
\andy{FockJC}
\beq
\rho_{\rm th}^{\rm JC} = \frac{1}{\cal Z} \sum^\infty_{j=0}
    \exp \left[ - j \beta \hbar \omega
    \right] \vert j><j\vert ,
\label{eq:FockJC}
\eeq
and the condition $\mbox{Tr} \rho_{\rm th}^{\rm JC} =1$ yields
\andy{Znew}
\beq
{\cal Z} = \frac{1}{1-e^{- \beta \hbar \omega}}.
\label{eq:Znew}
\eeq
Notice the correspondence with the $N \rightarrow \infty$
limit of eqs.(\ref{eq:thinit})-(\ref{eq:Zfunc}),
and remember the importance of choosing the symmetrized space
${\cal H}_N$ for the spin case: As already stressed, the unsymmetrized space
${\cal H}_{\{N\}}$ would have given different expressions
in eqs.(\ref{eq:repth}) and~(\ref{eq:Zfunc}).

The initial maser state is characterized by the quantities
\andy{invalues}
\barr
<H_D^{\rm JC}>_I^{\rm th} & = & \hbar \omega \, \kappa_{\rm th},  \nonumber \\
<\delta H_D^{\rm JC} >_I^{\rm th} & = & \hbar \omega \, \kappa_{\rm th}
            e^{\beta\hbar\omega/2} ,
         \nonumber \\
\frac{<\delta H_D^{\rm JC} >_I^{\rm th}}{<H_D^{\rm JC} >_I^{\rm th}} & = &
            e^{\beta\hbar\omega/2} ,
\label{eq:invalues}
\earr
where $\kappa_{\rm th} = e^{-\beta\hbar\omega} /
\left( 1-e^{-\beta\hbar\omega} \right)$
is the number of boson excitations in the initial thermal state.
This is identical to eq.(\ref{eq:thlim1}).
It is not difficult to prove that
\andy{outval}
\barr
<H_D^{\rm JC}>_F^{\rm th} & = & \hbar \omega \left(
            \kappa_{\rm th} + \overline{\kappa} \right) ,
                    \nonumber \\
<\delta H_D^{\rm JC}>_F^{\rm th} & = & \hbar
            \omega \, \kappa_{\rm th}
            \sqrt{ 1 + \left( 2 \overline{\kappa} + 1 \right)
            \left( e^{\beta\hbar\omega}-1 \right)
            + \overline{\kappa} \left(
            e^{\beta\hbar\omega}-1 \right)^2 } \ ,
  \label{eq:outval}  \\
\frac{<\delta H_D^{\rm JC}>_F^{\rm th}}{<H_D^{\rm JC}>_F^{\rm th}} &
            = & \frac{\sqrt{1 + \left(
            2 \overline{\kappa} + 1 \right)
            \left( e^{\beta\hbar\omega}-1 \right)
            + \overline{\kappa} \left(e^{\beta\hbar\omega}-1 \right)^2 }}{1+
            \overline{\kappa} \left(e^{\beta\hbar\omega}-1 \right) } \ .
\nonumber
\earr
Once again, the correspondence with
eq.(\ref{eq:thlim}) is perfect.
Of course, the values of eq.(\ref{eq:JCgro}) are recovered for
$\theta =0 (\beta = \infty)$.

Observe also that, in agreement with eq.(\ref{eq:thfluclim}),
\andy{thfluclimJC}
\beq
\frac{<\delta H_D^{\rm JC}>_F^{\rm th}}{<H_D^{\rm JC}>_F^{\rm th}} \left\{
  \begin{array}{ll}
   \stackrel{\beta=0}{\longrightarrow} & 1  \\
   \stackrel{\beta=\infty}{\longrightarrow} & 1/\sqrt{\overline{\kappa}} .
  \end{array}
\right.
\label{eq:thfluclimJC}
\eeq

The calculation for the visibility of the interference pattern is
somewhat more involved, and is given in Appendix~C.
The result is
\andy{JCvis}
\beq
{\cal V}^{\rm JC}_{\rm th} =
 \exp \left[ - \left( \kappa_{\rm th} + \frac{1}{2} \right)
           \overline{\kappa} \right] ,
\label{eq:JCvis}
\eeq
and is identical to the value given in eq.(\ref{eq:thlim}).

\setcounter{equation}{0}
\section{Identifying the Hamiltonians}
\label{sec-Hlim}
\andy{Hlim}

{}From the complete correspondence between the physically interesting
quantities
calculated in the JC case and in the macroscopic ($N \rightarrow \infty$)
limit of the modified AgBr case, we may expect that there exists a macroscopic
limit of the modified AgBr Hamiltonian
(\ref{eq:H}), which reproduces the JC Hamiltonian
(\ref{eq:HJC}).
We have already seen that as far as the $S$-matrix is concerned,
the detailed structure of the potential $V$ does not play any role:
Only the integrated quantity $\int_{- \infty}^{\infty} V(x) dx = V_0 \delta$
has relevance.
Notice also that we have restricted our attention to the
${\cal P}_N$-invariant sector ${\cal H}_N$
of the total space ${\cal H}_{\{N\}}$, so that
we are mainly interested in
^^ ^^ global" quantities like the total number of spin flips, and
information like which spins are flipped and which are not is of no
importance, in particular in the macroscopic limit to be
considered.
Therefore, we can try to neglect the $x_n$-dependence of the potential
$V( \h{x} -x_n)$ from the beginning, in order to estabilish
a link between the two Hamiltonians
(\ref{eq:H}) and (\ref{eq:HJC}).

Here we shall consider two of the possible limiting procedures
for the modified AgBr Hamiltonian.
Since we have already estabilished the $N \rightarrow \infty$ limit
for the free Hamiltonian $H_D$ (see eq.(\ref{eq:sumup})),
let us concentrate our attention on the interaction Hamiltonian
\andy{Intonly}
\beq
H' = \sum_{n=1}^{N}  V(\h{x}- x_n)
  \sigma_{1}^{(n)} \exp \left( i \frac{\omega}{c}
  \sigma_{3}^{(n)} \h{x} \right) .
\label{eq:Intonly}
\eeq
One possibility to take the $N \rightarrow \infty$ limit
is to consider the case in which the
spins are all placed at the same position, say
\andy{sameplace}
\beq
x_n \equiv x_0 = 0, \qquad \forall n=1, \ldots , N.
\label{eq:sameplace}
\eeq
Another possibility is to consider a kind of average potential
over the positions of the scatterers $x_n$, and replace $V(\h{x}-x_n)$
with its average (say $\overline{V} (\h{x})$). In the
latter case, we are implicitly
assuming that all spins are distributed in a rather small region \cite{NaPa2}.

In either case we obtain
(writing $V$ for $\overline{V}$ in the latter case)
\andy{reH}
\barr
\! H' \! & = & V(\h{x}) \sum_{n=1}^{N}
  \sigma_{1}^{(n)} \exp \left( i \frac{\omega}{c}
  \sigma_{3}^{(n)} \h{x} \right)    \nonumber \\
   & = & V(\h{x}) \sum_{n=1}^{N}
  \left[ \sigma_{+}^{(n)} \exp \left( -i \frac{\omega}{c}
  \h{x} \right) + \sigma_{-}^{(n)} \exp \left( + i \frac{\omega}{c}
  \h{x} \right) \right]  \nonumber \\
   & = & V(\h{x}) \sqrt{N} \left[
         \exp \left( -i \frac{\omega}{c} \h{x} \right)
         \frac{1}{\sqrt{N}} \sum_{n=1}^{N} \sigma_{+}^{(n)}
         + \exp \left( + i \frac{\omega}{c} \h{x} \right)
         \frac{1}{\sqrt{N}} \sum_{n=1}^{N} \sigma_{-}^{(n)} \right] \! .
\label{eq:reH}
\earr
Once again, the condition $qN=$ finite, with
$\sqrt{q} \simeq V_0 \delta / \hbar c = \int V(y) dy / \hbar c$,
implies that the quantity $V(\h{x}) \sqrt{N} = u(\h{x})$
behaves well in the $N \rightarrow \infty$ limit.
Therefore, in the macroscopic limit, we can see that the modified AgBr
Hamiltonian is transformed into the JC Hamiltonian
(\ref{eq:HJC}).
In conclusion, the $N$-spin system behaves, in the $N \rightarrow \infty$
limit,
as a ^^ ^^ cavity", in which boson-like excitations (collective modes)
can be created, as a consequence of the interaction with the $Q$-particle.

We close this section with a remark: The AgBr and JC Hamiltonians have been
identified when
the detailed internal structure of the particle-spin interaction
and/or the spin locations are neglected
(see for instance eq.(\ref{eq:sameplace})).
However, this assumption is {\em not} fundamental
because, as we have seen in
Sec.~\ref{sec-masersys},
all the physically interesting quantities, such as energy,
energy fluctuation, visibility of the interference pattern and so on,
can be calculated by making use {\em only}
of the $S$-matrix, whose limit can be computed in full generality,
as seen in Sec.~\ref{sec-Smatrlim}.


\setcounter{equation}{0}
\section{An interpretation }
\label{sec-games}
\andy{games}

In the previous sections we have seen that there is a nice correspondence
between the $N \rightarrow \infty$ limit of the AgBr model and the
JC model. We have proven the correspondence of the $S$-matrices
and the Hamiltonians in
Secs.~\ref{sec-Smatrlim} and~~\ref{sec-Hlim}, respectively,
and have shown the identity of the final results
when the $D$-system is initially in the ground state and in a thermal state
in Secs.~\ref{sec-maserground} and \ref{sec-JCth}, respectively.

In the present section,
we wish to push further the correspondence between the two cases.
In order to do this, we shall introduce a suitable notation to denote
generalized and Glauber coherent states, and shall explicitly compute
the relevant evolutions. It turns out convenient, in the following, to suppress
the $Q$-particle states. Needless to say, these could be explicitly taken into
account, but at the price of making the notation cumbersome
and the formulae more
involved.
Therefore, in this section, we shall exclusively consider
the $D$-states. This can be accomplished via the
following expedient:
Let us consider the extreme situation in which the energy of the
$Q$-system is so large that the loss of energy due to the interaction
with the spin system can safely be neglected. That is, we assume that
$cp \gg \hbar \omega$.
Notice that we still keep
$H_D \neq 0$.

Under the above-mentioned condition,
if we take an initial ground state, the evolution of the total system may be
written as
\andy{zerodel}
\barr
S^{[N]} \vert p, 0>_N & \simeq & \vert p>
 \sum_{j=0}^N {N\choose j}^{1/2}
 \left( -i\sqrt{q} \right)^j \left( \sqrt{1-q} \right)^{N-j}
 \vert j>_N  \nonumber \\
    & \equiv & \vert p >\vert -i \sqrt{q} >_N ,
\label{eq:zerodel}
\earr
because the state $\vert p - j \frac{\hbar \omega}{c} , j>_N
\simeq \vert p , j>_N$ for small $j$, and the probability of losing a
large amount of energy (for large $j$) is very small
$(\simeq q^N)$ for small $q \simeq O(N^{-1})$.
Here the state $\vert -i \sqrt{q}>_N$ is a generalized
coherent state \cite{KMK},
and the $Q$-particle state, being factorized, can be neglected.
Incidentally, we stress the correspondence between this
case and the original Coleman-Hepp model
reviewed in Sec.~\ref{sec-agbrgen}.
We can write
\andy{vacN}
\beq
S^{[N]} \vert 0>_N \simeq \vert -i \sqrt{q} >_N
    \equiv \vert 0^\star >_N ,
\label{eq:vacN}
\eeq
where $\vert 0^\star >_N$ is the usual outgoing state of
scattering theory, and
represents here a new ^^ ^^ vacuum",
in a sense to be clarified later.
Analogously, for the JC case, from eq.(\ref{eq:SvacJC}) we get
\andy{vacJC}
\beq
S \vert 0> \simeq e^{- \overline{\kappa}/2}
   \sum_{j=0}^\infty \frac{(-i \sqrt{\overline{\kappa}} \, )^j}{\sqrt{j!}}
   \vert j> \equiv \vert -i \sqrt{\overline{\kappa}} > \equiv
   \vert 0^\star > ,
\label{eq:vacJC}
\eeq
where the coherent state $\vert -i \sqrt{\overline{\kappa}} > $
has been written
in the $z$-representation ($a \vert z> = z \vert z>$).
Obviously, the $N \rightarrow \infty, \; qN = \overline{n} =
\overline{\kappa}$ limit of the r.h.s.~of eq.(\ref{eq:zerodel}) yields
the coherent state of eq.(\ref{eq:vacJC}).
Once again, the action of the $S$-matrix on the vacuum $\vert 0>$
has the effect of generating a new vacuum.
In the same spirit, by making use of eq.(\ref{eq:KMKorigrel}),
we define
\andy{numberN}
\barr
S^{[N]} \vert m>_N
 & = & S^{[N]} \frac{ \left( \sqrt{N} \Sigma_+^{(N)} \right)^m }
   {\sqrt{m!} \prod_{k=0}^{m-1} \!
    \sqrt{ 1 - \frac{k}{N}}} \; \vert 0 >_N \,
   \nonumber \\
 & = & \frac{ \left( \sqrt{N} \Sigma_+^{\star(N)} \right)^m }
   {\sqrt{m!} \prod_{k=0}^{m-1} \!
    \sqrt{ 1 - \frac{k}{N}}} \; \vert 0^\star >_N \,
    \equiv \vert m^\star >_N ,
   \label{eq:numberN}     \\
  & & \qquad \sqrt{N} \Sigma_\pm^{\star(N)} \equiv
   S^{[N]} \sqrt{N} \Sigma_\pm^{(N)} S^{[N]\dagger} \nonumber ,
\earr
and, in the JC case,
\andy{numberJC}
\barr
S \vert m>
 & = & S \frac{ \left( a^{\dagger } \right)^m }{\sqrt{m!}}
   \vert 0 > \,
   \nonumber \\
 & = & \frac{ \left( a^{\star \dagger } \right)^m }{\sqrt{m!}}
   \vert 0^\star > \,
    \equiv \vert m^\star >,
\label{eq:numberJC}    \\
  & &   \qquad a^\star \equiv
   S a S^\dagger = a + i \sqrt{\overline{\kappa}}. \nonumber
\earr

In the thermal case, the evolution is given by
\andy{thevol}
\barr
\rho_{\rm th}^F =
S^{[N]} \rho_{\rm th} S^{[N]\dagger}
  & \simeq & \frac{1}{Z} \exp \left[ -\beta \frac{\hbar \omega}{2}
  S^{[N]} \sum_{n=1}^{N}  \left( 1+\sigma_{3}^{(n)} \right) S^{[N]\dagger}
  \right] ,  \nonumber \\
  & = & \frac{1}{Z} \exp \left[ -\beta \frac{\hbar \omega}{2}
  \sum_{n=1}^{N}  \left( 1+\sigma_{3}^{\star(n)} \right)
  \right] ,
\label{eq:thevol}
\earr
where the notation is the same as in
eq.(\ref{eq:partev}).
Analogously, one gets
\andy{thevJC}
\beq
S \rho_{\rm th}^{\rm JC} S^\dagger
   \simeq \frac{1}{\cal Z} \exp \left[ -\beta \hbar \omega
                a^{\star \dagger} a^\star \right] .
\label{eq:thevJC}
\eeq
It is easy to prove that this is the same quantity
obtained in the $N \to \infty$ limit from eq.(\ref{eq:thevol}).

Summarizing, if the energy change of $Q$ is neglected,
we understand the following correspondence
in the $N \rightarrow \infty, qN =\overline{n}< \infty$ limit,
\andy{finalsum}
\barr
\sqrt{N} \Sigma_{+}^{(N)}
   & \longrightarrow & a^\dagger, \nonumber \\
\sqrt{N} \Sigma_{-}^{(N)}
   & \longrightarrow & a, \nonumber \\
\frac{N}{2} \left( {\bf 1}^{(N)} + \Sigma_{3}^{(N)} \right)
   & \longrightarrow & {\cal N} =
      a^\dagger a,    \nonumber \\
S^{[N]}
   & \longrightarrow & S,    \label{eq:finalsum} \\
\vert 0 >_N & \longrightarrow & \vert 0 > , \nonumber \\
\vert m >_N & \longrightarrow & \vert m > ,  \nonumber \\
\rho_{\rm th} & \longrightarrow & \rho_{\rm th}^{\rm JC}  ,  \nonumber
\earr
and analogously for the $\,^\star$ states and operators.
We are now ready to put forward an interpretation of the results obtained in
the previous sections. First,
notice that in this paper we have not considered ^^ ^^ decoherence"
effects \cite{NaPa2,MN}, namely we have not tried to understand
why and how the density matrix of the $Q+D$ systems evolves
from a pure to a mixed state:
We have simply
introduced a solvable dynamical model describing the interaction between a
particle and a ^^ ^^ detector", without fully addressing the
problem of the loss of quantum coherence.
{}From the measurement-theoretical point of view,
the interest of the present model lies in the
appearance of a superselection-rule space in the
$qN = \overline{n} =
\overline{\kappa} \rightarrow \infty$ limit.
The phenomenon is well known in
the many-Hilbert-space theory \cite{MN}, where
the macroscopic apparatus (detector) is described
by means of a unitary inequivalent representation.

In order to understand the above-mentioned point,
observe first that the visibility of the interference pattern
(\ref{eq:alllim}) and~(\ref{eq:JCgro})
disappears
in the $\overline{n} = \overline{\kappa} \rightarrow \infty$ limit
as the two ^^ ^^ vacua" become orthogonal:
\andy{ineqrep}
\beq
{\cal V} =
<0 \vert S \vert 0> \simeq <0 \vert 0^\star > = e^{- \overline{\kappa} /2 }
\, \stackrel{\overline{\kappa} \rightarrow \infty}{\longrightarrow} \,
0 .
\label{eq:ineqrep}
\eeq
Most of the analyses previously performed on the AgBr Hamiltonian
have dealt with this situation.

Moreover, it is very
interesting to rewrite the visibility in the thermal case
(\ref{eq:thlim}) and~(\ref{eq:JCvis}) as
\andy{visineq}
\barr
{\cal V}^{\rm th} = \mbox{Tr} ( \rho_{\rm th} S )
& \simeq &
\sum_j \frac{e^{- j \beta \hbar \omega}}{\cal Z}
   <j \vert e^{-i \sqrt{\overline{\kappa}}(a^\dagger + a)} \vert j>
  \nonumber \\
& = &
\sum_j \frac{e^{- j \beta \hbar \omega}}{\cal Z}
   <0 \vert \frac{a^j}{\sqrt{j!}} \frac{ \left( a^{\dagger} -i
\sqrt{\overline{\kappa}}\right)^j}{\sqrt{j!}}
e^{-i \sqrt{\overline{\kappa}}(a^\dagger + a)} \vert 0>
  \nonumber \\
 & = & \sum_j \frac{e^{- j \beta \hbar \omega}}{\cal Z}
   <0 \vert \frac{a^j \left( a^{\star \dagger} \right)^j}{j!} \vert 0^\star >
  \nonumber \\
 & = & \sum_j \frac{e^{- j \beta \hbar \omega}}{\cal Z}
   <j \vert j^\star >  =
     e^{- \left(\kappa_{\rm th} + \frac{1}{2} \right) \overline{\kappa}}
\, \stackrel{\overline{\kappa} \rightarrow \infty}{\longrightarrow} \, 0.
\label{eq:visineq}
\earr
One clearly sees that interference disappears as the two basis
$\{\vert j >\}$ and $\{\vert j^\star >\}$ become
{\em orthogonal} to each other.
In this sense, we may say that the
$N\rightarrow \infty$, $qN = \overline{n}= \overline{\kappa} =$ finite limit
considered in this paper
^^ ^^ foreruns" the appearance of a {\em superselection-rule} space.

The visibility is often considered as a physical quantity able to
characterize the loss of coherence between the two interfering branch waves
(^^ ^^ collapse of the wave function"). This is not always correct: Indeed,
even though a loss of quantum coherence implies a loss of
interference, the opposite is not necessarily true, because the
interference pattern may vanish even though the total $Q+D$ system is
still in a pure state \cite{SW}.
In the case described in the present paper, all evolutions are described
by $S$-matrices and are therefore strictly {\em unitary}.
If the initial state is a pure state,
the final state remains pure, and in this sense
quantum coherence is always preserved.

Nevertheless, one may safely regard the above result as a first
step towards the loss of quantum coherence (^^ ^^ collapse"), because
the $D$-system, being macroscopic, undergoes internal motions
that tend to destroy the delicate coherence between its elementary
constituents. In the AgBr model considered, for example,
we have neglected the presence of interactions between the molecules,
as well as their positions (the $x_n$'s play the role of simple
parameters, and not of dynamical variables). All these additional effects,
if taken into account, would have randomized the process and
provoked decoherence, so that statistically, after many repetitions
of the ^^ ^^ experiment", phase-correlation effects
would have been washed out. In this statistical sense one can state
that if all additional randomization processes had been taken into account
the ^^ ^^ collapse of the wave function" would have occurred.

\setcounter{equation}{0}
\section{Additional remarks }
\label{sec-concl }
\andy{concl }

We have studied the interaction between an ultrarelativistic
particle $Q$ and a ^^ ^^ detector" $D$, schematized as a linear
array of two-level systems (^^ ^^ AgBr molecules"), that can be
excited (dissociated)
as a consequence of the interaction. We have seen that if the original
AgBr model is suitably modified, it is possible to take into account
energy-exchange processes between $Q$ and $D$: This is physically appealing,
because the state of a detector should show trace
of the passage of the particle, also from an energetic viewpoint.
We have computed the macroscopic limit of this system
and stressed a correspondence
with the Jaynes-Cummings model.

As mentioned in the previous section, the examples considered in this paper
are particularly interesting from the point of view of quantum measurements.
We have seen that the visibility has a remarkable behaviour
in all the cases considered, and in particular in the
macroscopic limit.
Notice, that while $\overline{n}=\overline{\kappa}$ represent the strength
of the interaction between $Q$ and $D$, $n_{\rm th}$ and
$\kappa_{\rm th}$ express the presence of (thermal) noise.
Obviously, from eqs.(\ref{eq:alllim}) and (\ref{eq:thlim})
(or alternatively, from eqs.(\ref{eq:JCgro}) and (\ref{eq:JCvis})),
the visibility disappears in both cases as
$\overline{n}=\overline{\kappa} \rightarrow \infty$:
This means, in a certain sense, that
the macrosystem ^^ ^^ works better" as a ^^ ^^ measuring system",
as the strength of the interaction
between $Q$ and $D$ increases.
On the other hand, as was to be expected,
as soon as the $Q$ and $D$ systems are dynamically
coupled ($\overline{n}=\overline{\kappa} \neq 0$),
the visibility
tends to vanish more quickly if the detector
is initially in a thermal state rather than in the ground state.

Notice also that the visibility vanishes quickly (exponentially) as the
temperature increases. If we consider, within the limits stressed
at the end of last section, the visibility as a physical quantity able to
characterize the loss of coherence between the two interfering branch waves
of the object system (^^ ^^ collapse of the wave function"), we realize
that the $Q$-system {\em decoheres} more as the temperature of the $D$
system increases.
In the above-mentioned sense, one could speak of {\em imperfect
measurements}: The visibility plays the role of a parameter that controls how
^^ ^^ effective" a measurement of the $Q$-particle trajectory is.
The value ${\cal V} = 1$ ($\overline{n}=\overline{\kappa} = 0$),
signifies absence of interaction between $Q$ and $D$: The $Q$-system
does not ^^ ^^ see" the detector and behaves as a ^^ ^^ wave".
Interference between the two branch waves is complete.
On the other hand, the value ${\cal V} = 0$
represents a ^^ ^^ particle" behaviour, and a total loss of interference.
Notice that the latter situation can be achieved if
$\overline{n}=\overline{\kappa} \rightarrow \infty$ or, alternatively,
when $D$ is initially in a thermal state, if
$\overline{n}=\overline{\kappa} \neq 0$ and
$n_{\rm th} = \kappa_{\rm th} \rightarrow \infty$ : This simply
means a nonvanishing interaction between $Q$ and a $D$-system
that is initially at $\infty$ temperature.
The intermediate cases $0 < {\cal V} < 1$ represent imperfect
measurements, after which the branch waves of the $Q$-system
are still able to interfere, at least to a partial extent.

We stress that the problem of decoherence and imperfect measurements
is certainly much more delicate than implied by the above discussion.
In particular, notice that the off-diagonal terms (with respect to $Q$)
of the total ($Q+D$) density matrix have not been shown to vanish,
in the cases considered in the present paper, so that, strictly speaking,
the problem of decoherence has not been addressed in its full generality.
More careful investigation is required on this point.

It is also interesting to comment on a remark put forward by
Busch, Lahti and Mittelstaedt \cite{Busch} about the occurrence of
nonseparable Hilbert spaces when (continuous) superselection rules
appear in the description of macroscopic apparatuses.
It seems to us that there are cases (and the model discussed in this
paper provides an example) in which physics itself
^^ ^^ suggests" which limit and space
are more suitable to describe the situation
investigated: In the AgBr system, one could have considered
other possible situations, such as, for instance,
the space ${\cal H}_{\{N\}}$ or the $N \rightarrow \infty$
limit without keeping the quantity $qN$ finite.
We have already observed that
$\dim {\cal H}_{\{N\}} = 2^N$,
so that in the
$N \rightarrow \infty$ limit the space
${\cal H}_{\{N\}}$ is nonseparable.
On the other hand,
$\dim {\cal H}_{N} = N+1$, and ${\cal H}_{N}$
tends to a separable Hilbert space:
In fact, the $qN$-finite case investigated in this paper
turns out to be equivalent to a maser system, that is describable
in Fock space.
It seems therefore that the requirement of physically reasonable
conditions (such as {\em finite} energy exchange between $Q$ and $D$ and
restriction to {\em symmetrized} $D$-states)
lead, in the thermodynamical limit,
to a {\em separable} Hilbert space,
and therefore the emergence of nonseparability
is not a necessary consequence of the appearance of superselection
rules.

The model discussed in this paper has proven to be a very fertile
example for discussions on quantum measurements.
Even though the argument remains open, in particular on the
problem of decoherence, we hope to have convincingly shown that a
quantum mechanical measurement process can be treated within
quantum mechanics, and one need not postulate a ^^ ^^ classical"
behavior for the measuring apparatus.
\vspace*{.5cm}

{\bf Acknowledgements} \vspace{.3cm}

The authors acknowledge interesting remarks and suggestions
by Profs.~S. Kudaka,
S.~Matsumoto,
M.~Namiki,
K.~Niizeki,
H.~Rauch,
J.~Summhammer,
S.~Takagi,
M.~Villani,
N.~Yamada
and in particular by Prof.~K.~Kakazu.
They thank the University of the Ryukyus Foundation and
the Italian National Institute for Nuclear Physics (INFN) for the financial
support. This work was also partially supported by the Grant-in-Aid
for Scientific Research of the Ministry of Education, Science and Culture,
Japan (03854017) and by Italian Research National Council
(CNR) under the bilateral
projects Italy-Japan n.91.00184.CT02 and n.92.00956.CT02.



\renewcommand{\thesection}{\Alph{section}}
\setcounter{section}{1}
\setcounter{equation}{0}
\section*{Appendix A}
\label{sec-appA}
\andy{appA}

     Let us first derive a general formula for the visibility ${\cal V}$.  We
consider a typical Young-type experiment displaying the interference pattern
(the treatment of a neutron-interferometric type experiment is analogous).
Let $\vert \psi_1 >$ and $\vert \psi_2 >$ be the two branch waves of
the initial $Q$-system and
$\vert j>$ a complete orthonormal set of the $D$-system
and assume that the evolution of the total system is described by the
$S$-matrix whose action is given by
\andy{Sonpsi}
\barr
S \vert \psi_1 >\vert j > & = &\vert \psi_1 >\vert j >,    \nonumber \\
S \vert \psi_2 >\vert j > & = & \sum_k C_{jk} \vert \psi'_{2k} >\vert k>.
\label{eq:Sonpsi}
\earr
We assumed that only $\vert \psi_2 >$ interacts with the detector $D$.

     Let, without loss of generality, the $D$-system be initially in
one of the above $\vert j >$ states, say $\vert n>$.  The final
state of the total system is described by the density matrix
\andy{finaldensity}
\barr
\rho_{\rm tot}^F
  & = & S\rho_{\rm tot}^IS^\dagger  \nonumber \\
  & = & S\Bigl[\left( \vert\psi_1><\psi_1\vert+\vert\psi_2><\psi_2\vert
      +\vert\psi_1><\psi_2\vert + \mbox{h.c.}
       \right) \otimes\vert n><n\vert\Bigr]S^\dagger
                                           \nonumber \\
  & = & \vert\psi_1><\psi_1\vert\otimes\vert n><n\vert+
        \sum_{j,k}C_{nj}C_{nk}^*\vert\psi'_{2j}><\psi'_{2k}\vert
                                                      \otimes\vert j><k\vert
                                            \nonumber \\
  &   &
        + \sum_jC_{nj}^*\vert\psi_1><\psi'_{2j}\vert
                                    \otimes\vert n><j\vert + \mbox{h.c.}.
\label{eq:finaldensity}
\earr
Therefore the probability of observing the particle
after the interaction, say at $\rz$, is given by
\andy{Pfind}
\barr
P(\rz)
  &=& \mbox{Tr}\Bigl[
      \vert\rz><\rz\vert\rho_{\rm tot}^{\rm F}\Bigr]         \nonumber \\
  &=&\vert\psi_1(\rz)\vert^2+\vert\psi_2(\rz)\vert^2
      +2\Re\Bigl[C_{nn}^*\psi_1(\rz)\psi'{}^*_{2n}(\rz)\Bigr],
\label{eq:Pfind}
\earr
where the trace is taken over both the $Q$ and $D$ states,
$\psi_1(\rz)=<\rz\vert\psi_1>$ is the branch-wave function of
the particle, and so on.  (Notice that the $S$-matrix, which is
responsible only for the interaction between the
$Q$ and $D$ systems, does not contain $\rz$ as a dynamical variable.)
The position $\rz$ corresponds to the location of the
particle's spot on the screen.
To simplify the discussion, we assume that the wave functions after
the interaction with the $D$-system can be well approximated by plane waves.
Then we understand that the first two terms in (\ref{eq:Pfind}) no longer
depend on $\rz$ and the interference pattern is produced by the last
term.  By assuming $\vert\psi_1\vert^2=\vert\psi_2\vert^2
=\vert\psi'_{2n}\vert^2$, the visibility
$\cal V$ is given by
\andy{visibility}
\beq
{\cal V}={P_{\rm MAX}-P_{\rm min}\over P_{\rm MAX}+P_{\rm min}}
        =\vert C_{nn}\vert,
\label{eq:visisbility}
\eeq
where $_{\rm MAX}$ and $_{\rm min}$ are relative
to the screen coordinate $\rz$.
Notice that $C_{nn}$ can also be written as
\andy{Cnn}
\beq
C_{nn}=\mbox{Tr}\Bigl[\vert\rz><\rz\vert\otimes\vert n><n\vert\,S\Bigr],
\label{eq:Cnn}
\eeq
or, if we suppress the $Q$-states,
\andy{Cnnbis}
\beq
C_{nn}=\mbox{Tr}\Bigl[ \vert n><n\vert\,S\Bigr]
= < n \vert S \vert n>.
\label{eq:Cnnbis}
\eeq
This is the formula used in eqs.(\ref{eq:vis}), (\ref{eq:visbis})
and (\ref{eq:JCgro}).

     Similarly, if the initial $D$-system is described by a density matrix
\andy{Cnnss}
\beq
\rho_D^I =\sum_np_n\vert n><n\vert,
\label{eq:Cnnss}
\eeq
the probability of
finding the particle at $\rz$ is calculated as
\andy{Pfind2}
\barr
P(\rz)
  &=& \mbox{Tr}\Bigl[
      \vert\rz><\rz\vert\rho_{\rm tot}^F\Bigr]         \nonumber \\
  &=&\vert\psi_1(\rz)\vert^2
      +\vert\psi_2(\rz)\vert^2
      +2\Re\Bigl[\sum_np_nC_{nn}^*\,\psi_1(\rz)\psi'{}^*_{2n}(\rz)\Bigr],
\label{eq:Pfind2}
\earr
which yields
\andy{visibility2}
\beq
{\cal V}=\sum_np_n\vert C_{nn}\vert .
\label{eq:visibility2}
\eeq
Thus we find that the visibility $\cal V$ is simply given by
\andy{visibility3}
\beq
\mbox{Tr}\Bigl[\vert\rz><\rz\vert\otimes\rho_D^I\,S\Bigr]
\label{eq:visibility3}
\eeq
or, if we suppress the $Q$-state, by
\andy{visibility4}
\beq
\mbox{Tr}\Bigl[\rho_D^I\,S\Bigr].
\label{eq:visibility4}
\eeq
This is the formula used in eqs.(\ref{eq:visN})
and (\ref{eq:JCvis}).

     Let us proceed to the explicit calculation of $\cal V^{\rm th}$ when the
initial $D$-state $\rho_D^I$ and the $S$-matrix are given by
$\rho_{\rm th}$ (\ref{eq:repth}) and $S^{[N]}$ (\ref{eq:Smatrbis}),
respectively.

First observe that the $S$-matrix in
eq.(\ref{eq:Smatrbis}) is expressed as
\andy{Smatrag}
\barr
S^{[N]} & = &
\exp \left( -i\alpha\left[ N
\Sigma^{(N)}_+ (\omega) + N \Sigma^{(N)}_- (\omega) \right]\right) ,
\label{eq:Smatrag}  \\
 & & \qquad
  \Sigma^{(N)}_\pm (\omega) \equiv
\frac{1}{N} \sum_{n=1}^{N}
\sigma_{\pm}^{(n)} \exp \left( \mp i \frac{\omega}{c} \h{x} \right) ,
 \nonumber
\earr
where $\alpha = V_0 \delta / \hbar c$.
Moreover,
$N \Sigma^{(N)}_\pm (\omega)$ and $N \Sigma^{(N)}_3$, which is
defined in  eq.(\ref{eq:Sigmaj}), satisfy the algebra
(\ref{eq:newalg}):
\andy{newalgbis}
\barr
\left[ N \Sigma^{(N)}_- (\omega),N \Sigma^{(N)}_+ (\omega) \right] & = &
-N \Sigma^{(N)}_3, \nonumber \\
\left[ N \Sigma^{(N)}_\pm (\omega),-N \Sigma^{(N)}_3 \right] & = &
\pm 2N \Sigma^{(N)}_\pm (\omega).
         \label{eq:newalgbis}
\earr
This allows us to rewrite $S^{[N]}$ as \cite{Bogol}
\andy{relbis}
\beq
S^{[N]} = e^{-i \tan\alpha \cdot N \Sigma^{(N)}_+(\omega)}
        e^{-\ln\cos\alpha \cdot N \Sigma^{(N)}_3}
        e^{-i \tan\alpha \cdot N \Sigma^{(N)}_-(\omega)}.
     \label{eq:relbis}
\eeq
It is straightforward to obtain \cite{NaPa2} \cite{KMK}
\andy{KMKrel}
\barr
N \Sigma^{(N)}_+ (\omega) \vert p,n>_N & = &
    \sqrt{(N-n)(n+1)} \; \vert p - \frac{\hbar \omega}{c} , n+1>_N,
\nonumber \\
N \Sigma^{(N)}_- (\omega) \vert p,n>_N & = &
    \sqrt{(N-n+1)n} \; \vert p + \frac{\hbar \omega}{c} , n-1>_N,
         \label{eq:KMKrel}  \\
N \Sigma^{(N)}_3 \vert p,n>_N & = &
    (2n-N) \; \vert p , n>_N,
\nonumber
\earr
so that
\andy{compex}
\barr
e^{-i \tan\alpha \cdot N \Sigma^{(N)}_-(\omega)} \vert p,n>_N
  & = & \sum_{k=0}^{\infty} \frac{ (-i \tan \alpha)^k}{k!}
        \left[ N \Sigma^{(N)}_-(\omega) \right]^k \vert p,n>_N
\nonumber \\
& = & \sum_{k=0}^{n} \frac{ (-i \tan \alpha)^k}{k!}
  \sqrt{ \frac{(N-n+k)!n!}{(N-n)!(n-k)!}}
  \; \vert p + k \frac{\hbar \omega}{c} ,n-k>_N , \nonumber \\
     \label{eq:compex}
\earr
and similarly
\andy{compex2}
\beq
\,_N<p,n \vert
e^{-i \tan\alpha \cdot N \Sigma^{(N)}_+(\omega)} =
  \sum_{k=0}^{n} \frac{ (-i \tan \alpha)^k}{k!}
  \sqrt{ \frac{(N-n+k)!n!}{(N-n)!(n-k)!}} \;
  \;_N \! < p + k \frac{\hbar \omega}{c} ,n-k \vert .
     \label{eq:compex2}
\eeq
Therefore
\andy{matel}
\barr
\,_N< p, n \vert S^{[N]} \vert p, n >_N
 & = &    \sum_{k=0}^{n} (-1)^k \tan^{2k} \! \alpha \,
      {N-n+k\choose k}{n\choose k} (\cos \alpha
      )^{N-2(n-k)}
    \nonumber \\
& = &  \cos^N \! \alpha
     \left( \frac{1}{\cos^2 \! \alpha} \right)^n
     \frac{1}{(N-n)!} \frac{d^{N-n}}{dx^{N-n}} \left.
     x^{N-n} (1+x)^n \right\vert_{x= -\sin^2 \alpha} .
    \nonumber \\
  \label{eq:matel}
\earr
Since $p_n$, in eq.(\ref{eq:visibility2}), is given by
$\exp (-n \beta \hbar \omega)/Z$,
the problem is reduced to evaluate
($g \equiv \exp (- \beta \hbar \omega )/ \cos^2 \alpha$)
\andy{sumk}
\beq
\sum_{n=0}^{N}
     \frac{g^n}{(N-n)!} \frac{d^{N-n}}{dx^{N-n}}
     x^{N-n} (1+x)^n ,
  \label{eq:sumk}
\eeq
which can be expressed as a complex $z$-integration
around $z=x$
\andy{compint}
\barr
 \sum_{n=0}^{N}
     \frac{g^n}{2 \pi i } \oint \frac{z^{N-n} (1+z)^n}{(z-x)^{N-n+1}} dz
   & = & \frac{1}{2 \pi i }
     \oint \frac{z^{N}}{(z-x)^{N+1}} \sum_{n=0}^{N}
     \left( \frac{(1+z)(z-x)g}{z} \right)^n dz
 \nonumber \\
 & = & \frac{1}{2 \pi i }
     \oint \frac{z^{N}}{(z-x)^{N+1}}
     \frac{1}{1- \frac{(1+z)(z-x)g}{z}} dz .
  \label{eq:compint}
\earr
In the last equality, an analytic term around $z=x$ has been omitted
because its contribution vanishes.
Changing the integration variable from $z$ to $t=(z-x)/z$, the last expression
becomes
\andy{cirint}
\beq
\frac{1}{2 \pi i }
     \oint \frac{dt}{t^{N+1}} f(t),
  \label{eq:cirint}
\eeq
where
\andy{cirint2}
\beq
f(t) = \frac{1}{1-t(1+(1+x)g)+gt^2 } ,
  \label{eq:cirint2}
\eeq
and the integration contour includes $t=0$. By writing $f(t)$ as
\andy{foft}
\beq
f(t) = \frac{1}{g(t_+ - t_-)}
     \left( \frac{1}{t - t_+} - \frac{1}{t- t_-} \right) ,
  \label{eq:foft}
\eeq
where
$t_\pm$ are the two roots of the denominator of $f(t)$
\andy{denroots}
\beq
t_\pm = \frac{1 + (1+x)g \pm \sqrt{(1 + (1+x)g)^2 -4g}}{2g} ,
  \label{eq:denroots}
\eeq
the integral in eq.(\ref{eq:cirint}) is readily evaluated to yield
\andy{intresu}
\beq
- \frac{1}{g(t_+ - t_-)}
     \left( \frac{1}{t_+^{N+1}} - \frac{1}{t_-^{N+1}} \right) .
  \label{eq:intresu}
\eeq
By taking into account the proper normalization factor, given by
the inverse of eq.(\ref{eq:Zfunc}), and evaluating eq.(\ref{eq:intresu})
at $x= -\sin^2 \alpha$, we finally arrive at the desired result
eq.(\ref{eq:visN}).

\addtocounter{section}{1}
\setcounter{equation}{0}
\section*{Appendix B}
\label{sec-appB}
\andy{appB}

We shall prove, following Ref.~\cite{contrac}, that the operators
$N^{-1/2} \sum_{n=1}^{N} \sigma_{\pm}^{(n)} =
N^{1/2} \Sigma_{\pm}^{(N)} $ and
$(1/2) \sum_{n=1}^{N} \left( 1+\sigma_{3}^{(n)} \right)
= (N/2) \left( {\bf 1}^{(N)} + \Sigma_{3}^{(N)} \right)$
obey, in the $N \rightarrow \infty$ limit,
the commutation relations for
$a, a^\dagger$ and ${\cal N} = a^\dagger a$.

We start from the generators of $SU(2)$ given in
eq.(\ref{eq:newalg}), and perform the following change of basis:
\andy{newbasis}
\beq
\left( \begin{array}{c}
h_+ \\
h_- \\
h_3 \\
1
\end{array} \right) =
\left( \begin{array}{cclc}
N^{-1/2} & & & \\
 & N^{-1/2} & & \\
 & & 1 & N/2 \\
 & & & 1
\end{array} \right)
\left( \begin{array}{c}
N \Sigma^{(N)}_+  \\
N \Sigma^{(N)}_-  \\
N \Sigma^{(N)}_3/2  \\
{\bf 1}^{(N)}
\end{array} \right) .
         \label{eq:newbasis}
\eeq
The commutation properties for ${\bf h}, 1$ are
\andy{newcomm}
\barr
\left[ h_3 , h_\pm \right] & = & \pm h_\pm ,
  \nonumber \\
\left[ h_- , h_+ \right] & = & 1 - \frac{2}{N} h_3 ,
  \nonumber \\
\left[ {\bf h} , 1 \right] & = & 0 ,
         \label{eq:newcomm}
\earr
and yield, in the $N \to \infty$ limit,
the standard boson commutation relations.
In conclusion,
\andy{sumup22}
\barr
h_+ = \frac{1}{\sqrt{N}} \sum_{n=1}^{N} \sigma_{+}^{(n)}
 = \sqrt{N} \Sigma_{+}^{(N)}
   & \stackrel{N \rightarrow \infty}{\longrightarrow} & a^\dagger, \nonumber \\
h_- = \frac{1}{\sqrt{N}} \sum_{n=1}^{N} \sigma_{-}^{(n)}
 = \sqrt{N} \Sigma_{-}^{(N)}
& \stackrel{N \rightarrow \infty}{\longrightarrow} & a,
 \label{eq:sumup22}    \\
h_3 = \frac{1}{2} \sum_{n=1}^{N} \left( 1 + \sigma_{3}^{(n)} \right)
 = \frac{N}{2} \left( {\bf 1}^{(N)} + \Sigma_{3}^{(N)} \right)
   & \stackrel{N \rightarrow \infty}{\longrightarrow} & {\cal N}
   \equiv a^\dagger a .
    \nonumber
\earr

\addtocounter{section}{1}
\setcounter{equation}{0}
\section*{Appendix C}
\label{sec-appC}
\andy{appC}

Here we compute the visibility of the interference pattern when
the cavity is initially in a thermal state.

The $S$-matrix for the maser case is given by
eq.(\ref{eq:SJC}) and can be rewritten as
\andy{SJC1}
\beq
S = \exp (- \overline{\kappa} /2)
\exp \left[ -i \sqrt{\overline{\kappa}}
         a^\dagger \exp \left( -i \frac{\omega}{c} \h{x} \right) \right]
         \exp \left[ -i \sqrt{\overline{\kappa}}
         a \exp \left( i \frac{\omega}{c} \h{x} \right)
         \right] ,
  \label{eq:SJC1}
\eeq
where $\overline{\kappa}$ was defined in eq.(\ref{eq:SvacJC}).
The initial thermal state $\rho^{\rm JC}_{\rm th}$
(eq.(\ref{eq:thinitJC})) has the following $P$-representation
\cite{Helstrom} in terms of coherent states
\andy{cohrep}
\beq
\rho^{\rm JC}_{\rm th} = \int \frac{d^2 \beta}{\pi \kappa_{\rm th}}
       \vert \beta > e^{- \vert \beta \vert^2 / \kappa_{\rm th}}
       < \beta \vert ,
  \label{eq:cohrep}
\eeq
where $a \vert \beta > = \beta \vert \beta >$ and
$\kappa_{\rm th}$ was defined after eq.(\ref{eq:invalues}).
By the same reasoning explained at the beginning of Appendix A,
we understand that the visibility is given by
\andy{vvisJJC}
\beq
{\cal V}^{\rm JC}_{\rm th} =
 \mbox{Tr}_D < p \vert \rho^{\rm JC}_{\rm th} S \vert p > ,
  \label{eq:vvisJJC}
\eeq
and is explicitly computed as
\andy{vviscomp}
\barr
{\cal V}^{\rm JC}_{\rm th} & = &
   \int \frac{d^2 \beta}{\pi \kappa_{\rm th}}
       < p, \beta \vert S \vert p, \beta >
       e^{- \vert \beta \vert^2 / \kappa_{\rm th}}
    \nonumber \\
& = &
   \exp (- \overline{\kappa} /2)
   \int \frac{d^2 \beta}{\pi \kappa_{\rm th}}
       < p \vert  \exp \left[ -i \sqrt{\overline{\kappa}}
         \beta^* \exp \left( -i \frac{\omega}{c} \h{x} \right) \right]
    \nonumber \\
&  & \qquad \qquad \times
 \exp \left[ -i \sqrt{\overline{\kappa}}
         \beta \exp \left( i \frac{\omega}{c} \h{x} \right)
         \right] \vert p >
       e^{- \vert \beta \vert^2 / \kappa_{\rm th}} .
  \label{eq:vviscomp}
\earr
Observe that
\andy{cohst}
\barr
\exp \left[ -i \sqrt{\overline{\kappa}}
         \beta \exp \left( i \frac{\omega}{c} \h{x} \right)
         \right] \vert p > & = &
    \sum_{n=0}^{\infty} \frac{(-i \sqrt{\overline{\kappa}}\beta)^n}{n!}
    e^{i n \omega \h{x} /c} \vert p >  \nonumber  \\
& = &
    \sum_{n=0}^{\infty} \frac{(-i \sqrt{\overline{\kappa}}\beta)^n}{n!}
    \vert p +  \frac{n \hbar \omega}{c} > ,
  \label{eq:cohst}
\earr
and similarly
\andy{cohst1}
\beq
< p \vert
\exp \left[ -i \sqrt{\overline{\kappa}}
         \beta^* \exp \left( - i \frac{\omega}{c} \h{x} \right)
         \right] =
    \sum_{n=0}^{\infty} \frac{(-i \sqrt{\overline{\kappa}}\beta^*)^n}{n!}
    < p +   \frac{n \hbar \omega}{c} \vert  .
  \label{eq:cohst1}
\eeq
Therefore eq.(\ref{eq:vviscomp}) becomes
\andy{vispart}
\beq
{\cal V}^{\rm JC}_{\rm th} =
   \exp (- \overline{\kappa} /2)
   \int \frac{d^2 \beta}{\pi \kappa_{\rm th}}
   \sum_{n=0}^{\infty} \frac{( \overline{\kappa} \vert \beta \vert^2)^n}{n!}
       e^{- \vert \beta \vert^2 / \kappa_{\rm th}} .
  \label{eq:vispart}
\eeq
The integration over $\beta$ is easily performed in two-dimensional
polar coordinates, and yields
\andy{visfin}
\barr
{\cal V}^{\rm JC}_{\rm th} & = &
   \exp (- \overline{\kappa} /2)
   \int_0^\infty \frac{d \vert \beta \vert^2}{\kappa_{\rm th}}
   \sum_{n=0}^{\infty} \frac{(- \overline{\kappa} \vert \beta \vert^2)^n}{n!}
       e^{- \vert \beta \vert^2 / \kappa_{\rm th}}
\nonumber \\
& = &
   \exp (- \overline{\kappa} /2)
   \sum_{n=0}^{\infty} \frac{1}{\kappa_{\rm th}}
\frac{(- \overline{\kappa})^n}{n!}
 \kappa_{\rm th}^{n+1} \Gamma (n+1)
\nonumber \\
& = &
   \exp (- \overline{\kappa} /2)
   \sum_{n=0}^{\infty}
\frac{( -\overline{\kappa} \kappa_{\rm th})^n}{n!}
\nonumber \\
& = &
   \exp \left[ - \left( \kappa_{\rm th} + \frac{1}{2} \right)
   \overline{\kappa} \right] .
  \label{eq:visfin}
\earr



\end{document}